\documentclass{v16nufact}

\usepackage{amsmath}
\usepackage{dsfont}
\usepackage[utf8]{inputenc}

\def\be{\begin{equation}}
\def\ee{\end{equation}}
\def\bea{\begin{eqnarray}}
\def\eea{\end{eqnarray}}

\newcommand{\Photo}{}

\begin{document}
\begin{flushright}
PSI-PR-16-15
\end{flushright}
\vspace*{4cm}
\title{Complementarity in lepton-flavour violating muon decay experiments}

\author{A.~Crivellin$^{1}$,
	S.~Davidson$^{2}$, 
        G.~M.~Pruna$^{1,\,}$\footnote{Corresponding author: \href{mailto:Giovanni-Marco.Pruna@psi.ch}{Giovanni-Marco.Pruna@psi.ch}.} and
	A.~Signer$^{1,3}$}

\address{\vspace{0.3cm}
${}^1$ Paul Scherrer Institut,\\
CH-5232 Villigen PSI, Switzerland \\
\vspace{0.3cm}
${}^2$ IPNL, CNRS/IN2P3, 4 rue E. Fermi,\\
69622 Villeurbanne cedex, France;\\ 
Universit\'e Lyon 1, Villeurbanne;
Universit\'e de Lyon, F-69622, Lyon, France\\
\vspace{0.3cm}
${}^3$ Physik-Institut, Universit\"at Z\"urich, \\
Winterthurerstrasse 190,
CH-8057 Z\"urich, Switzerland}

\maketitle\abstracts{ This note presents an analysis of
  lepton-flavour-violating muon decays within the framework of a
  low-energy effective field theory that contains higher-dimensional
  operators allowed by QED and QCD {symmetries}. The decay modes $\mu\to e \gamma$
  and $\mu\to 3e$ are investigated below the electroweak
  symmetry-breaking scale, down to energies at which such processes
  occur, \emph{i.e.}  the muon mass {scale}. The complete class of dimension-5
  and dimension-6 operators is studied systematically at the tree
  level, and one-loop contributions to the renormalisation group
  equations are fully taken into account. Current experimental limits
  are used to extract bounds on the Wilson coefficients of some of the
  operators and, ultimately, on the effective couplings at any energy
  level below the electroweak symmetry-breaking scale. Correlations
  between two couplings relevant to both processes illustrate the
  complementarity of searches planned for the MEG~II and Mu3e
  experiments.
\ }

\begin{boldmath}                                            %
\section{Introduction}
\label{sec:1}
\end{boldmath}                                              %
\noindent
This note presents a specific example of a correlation that occurs in
lepton-flavour-violating (LFV) muonic decays in the context of
effective field theories (EFTs). 

Whilst in the neutrino sector evidence for LFV is now established
beyond doubt~\cite{Fukuda:1998mi,Ahmad:2001an,Ahmad:2002jz}, the
absence of experimental hints of LFV in the charged lepton sector,
together with the smallness of the neutrino mass scale, indicate that
a very incisive flavour conservation mechanism is at work. Although
allowed in the Standard Model (SM) with right-handed neutrinos, the
branching ratios (BRs) of such transitions are suppressed by
$\left(m_\nu/M_W\right)^4$, making them too small to be observable in
any conceivable experiment. Consequently, any LFV production channel
or decay mode offers a promising benchmark against
which to search for physics beyond the SM.

Among charged LFV processes, muonic transition occurs in a relatively
clean experimental environment, to the point that the MEG experiment
has recently set  a stringent limit~\cite{TheMEG:2016wtm} on
${\rm{BR}}\left(\mu \to e\gamma\right)$. This  represents the
strongest existing bound on `forbidden' decays, while the SINDRUM
result~\cite{Bellgardt:1987du} obtained almost three decades ago is
still very competitive with regard to the current experimental status
of other sectors. The well-known outcomes of these experiments are: 
\begin{align}
{\rm Br}\left( \mu \to e \gamma \right)&\leq4.2 \times10^{-13}\,,\\
{\rm Br}\left(\mu \to 3e \right)&\leq1.0 \times10^{-12}.
\end{align}

Furthermore, there are good prospects for future MEG II and Mu3e
experiments. The former is expected~\cite{Baldini:2013ke} to reach a
limit of $4\times 10^{-14}$, while the latter might even achieve a
four-orders-of-magnitude
improvement~\cite{Blondel:2013ia,Perrevoort:2016nuv} on the
existing limit. All the aforementioned experiments are being carried
out at the Paul Scherrer Institut's experimental facilities. The
present analysis does not consider LFV transitions in a nuclear
environment (coherent and incoherent muon conversion in nuclei). See
Refs~\cite{Geib:2015unm}~and~\cite{Geib:2016atx} for extensive
treatments of this topic. 

From a theoretical perspective, LFV processes have been studied in
many specific extensions of the SM. In some cases the matching of such
extensions to a low-energy effective theory has also been
considered~\cite{Beneke:2015lba,Davidson:2016utf}. However, this
analysis follows a bottom-up approach in which effective interactions
are included in a low-energy Lagrangian~\cite{Kuno:1999jp} that
respects the $SU(3)_c$ and $U(1)_{EM}$ {gauge} symmetries. In exploiting the
Appelquist-Carazzone theorem~\cite{Appelquist:1974tg}, it is possible
to extend the QCD and QED Lagrangian\footnote{{Without the top quark field.}} with higher-dimensional operators
\begin{align}\label{eq:bw}
\mathcal{L}_{\rm eff} =\mathcal{L}_{\mathrm{QED+QCD}} + \frac{1}{\Lambda} \sum_{k} C_k^{(5)} Q_k^{(5)}
+ \frac{1}{\Lambda^2} \sum_{k} C_k^{(6)} Q_k^{(6)} +
\mathcal{O}\left(\frac{1}{\Lambda^3}\right)\, .
\end{align}
Here, $\Lambda$ is the ultraviolet (UV) completion energy scale,
which in this context is required not to exceed the electroweak
symmetry-breaking (EWSB) scale, where the SM dynamic degrees of
freedom and symmetries must be adequately
restored~\cite{Crivellin:2013hpa,Pruna:2014asa} and matched
with those of the low-energy theory.

Having established the theoretical background, the main focus is on
the interpretation of correlations between operators in the BRs of
both $\mu \to e \gamma$ and $\mu\to 3e$ at the muon mass energy scale
and beyond. Experimental limits are then applied to the parameter
space in a search for allowed regions.

{
The popular parametrisation of dipole and four-fermion LFV operators~\cite{deGouvea:2013zba}
\begin{align}\label{k-lag}
\mathcal{L}=\frac{m_{\mu}}{\left(k+1\right)\Lambda^2}\left(\bar{\mu}_R\sigma_{\mu\nu}e_L\right)F^{\mu\nu}+\frac{k}{\left(k+1\right)\Lambda^2}\left(\bar{\mu}_L\gamma_{\mu}e_L\right)\left(\bar{f}\gamma^\mu f\right),
\end{align} 
where $k$ is an \emph{ad hoc} parameter to be interpreted strictly at
the muon mass energy scale, allows to switch from a pure dipole
interaction ($k\sim 0$) to a pure four-fermion interaction ($k\sim
\infty$). Although this approach ensures a descriptive
phenomenological understanding of the contributions of different
operators to different observables, a more consistent theoretical
approach can be achieved without losing interpretive power.
}

The advantage of a systematic effective QFT approach lies in the fact
that it can be used to link phenomenological observables {at} different
energy scales unambiguously through the renormalisation-group
evolution (RGE) of the Wilson coefficients. In this regard, the RGE
between the muon mass energy scale and the EWSB scale is calculated at
the leading order (up to the one-loop level) in QED and QCD for any
operator contributing to LFV muon decays. This encompasses possible
mixing effects among operators, which in this study are taken into
account in a similar way to recent theoretical
works~\cite{Davidson:2016edt}. From this analysis, it is possible to
extract limits both on the Wilson coefficients defined at the
phenomenological energy scale and on the coefficients defined at the
UV matching scale.

This paper is organised as follows. Section~\ref{sec:2} introduces the
LFV effective Lagrangian, and in Section~\ref{sec:3} the observables
connected with the $\mu \to e\gamma$ and $\mu\to 3e$ searches are
briefly discussed. Section~\ref{sec:4} provides a brief
phenomenological analysis, and in Section~\ref{sec:5} conclusions are
drawn. Formulae relevant to the RGE of the Wilson coefficients are
provided in the appendix.

\vspace{1.2cm}
\begin{boldmath}                                            %
\section{LFV effective Lagrangian at the muon energy scale}%
\label{sec:2}
\end{boldmath}                                              %
\noindent
The Appelquist-Carazzone theorem~\cite{Appelquist:1974tg} is exploited
to construct an effective Lagrangians valid below the EWSB scale, with
higher-dimensional operators that respect the QCD $SU(3)_c$ and QED
$U(1)_{EM}$ symmetries. This allows for an interpretation of BSM
effects at high energy scales in terms of new, non-renormalisable
interactions at the low energy scale.

In this respect, all possible QCD and QED invariant operators relevant
to $\mu\to e$ transitions are considered up to dimension 6. These can
be arranged in the following effective Lagrangian with dimensionless
Wilson coefficients $C$ and the decoupling energy scale $M_W\ge
\Lambda \gg m_b$:
\begin{align}
\mathcal{L}_{\rm eff}&= \mathcal{L}_{\mathrm{QED+QCD}} +
\frac{1}{\Lambda^2}\left\{C_L^DO_L^D + \sum\limits_{f = q,\ell } 
{\left( {C_{ff}^{V\;LL}O_{ff}^{V\;LL} + C_{ff}^{V\;LR}O_{ff}^{V\;LR} + C_{ff}^{S\;LL}O_{ff}^{S\;LL}} \right)}\right.\nonumber \\
&\left. + \sum\limits_{f = q,\tau } {\left( {C_{ff}^{T\;LL}O_{ff}^{T\;LL} + C_{ff}^{S\;LR\;}O_{ff}^{S\;LR\;}} \right)}+ L \leftrightarrow R\right\} + \mathrm{H.c.},
\label{Leff}
\end{align}
where $q$ and $l$ specify that sums run over the quark and lepton
flavours, respectively. The explicit structure of the operators is
given by  
\begin{align}
\label{eq:magnetic}
O_L^{D} &= e \, m_\mu\left( \bar e{\sigma ^{\mu \nu }}{P_L}\mu\right) {F_{\mu \nu }}
,\\
O_{ff}^{V\;LL} &= \left(\bar e{\gamma ^\mu }{P_L}\mu\right) \left( \bar f{\gamma _\mu }{P_L}f\right)
,\\
O_{ff}^{V\;LR} &= \left(\bar e{\gamma ^\mu }{P_L}\mu\right) \left( \bar f{\gamma _\mu }{P_R}f\right)
,\\\label{eq:fourferm}
O_{ff}^{S\;LL} &= \left(\bar e{P_L}\mu\right) \left( \bar f{P_L}f\right)
,\\
O_{ff}^{S\;LR} &= \left(\bar e{P_L}\mu\right) \left( \bar f{P_R}f\right)
,\\
O_{ff}^{T\;LL} &= \left(\bar e{\sigma _{\mu \nu }}{P_L}\mu\right) \left( \bar f{\sigma ^{\mu \nu }}{P_L}f\right),
\label{Ogg}
\end{align}
and an analogous notation is assumed for cases in which the $L
\leftrightarrow R$ exchange is applied. In the previous equations, the
convention $P_{L/R}=\left(\mathds{1}\mp \gamma^5\right)/2$ is
understood. Apart from being multiplied by the QED coupling $e$, the
operator in Eq.\ref{eq:magnetic} is also rearranged into a dimension-6
operator with an appropriate normalisation factor $m_\mu$.  The reason
is that this operator is directly related to a dimension-6 operator in
the SMEFT~\cite{Buchmuller:1985jz,Grzadkowski:2010es}.

Direct comparison of Eq.~\ref{Leff} and Eq.~\ref{k-lag} reveals that
the latter assumes a tree-level correlation between independent
operators. This assumption is manifestly inconsistent when quantum
fluctuations are considered. Notably, an analysis of LFV transitions
in nuclei calls for a further dimension-7 operator relating to the
leading-order muon-electron-gluon interaction, which is generated by
threshold corrections induced by the heavy quark operators (see
Ref.~\cite{Cirigliano:2009bz} for details).

\vspace{1.2cm}
\begin{boldmath}                                            %
\section{Lepton-flavour-violating muonic observables}
\label{sec:3}
\end{boldmath}                                              %
\noindent
This section describes two of the most relevant LFV muon decay
processes, $\mu^+ \to e^+\gamma$ and $\mu^+\to e^+ e^- e^+$. Since the
following analysis does not include a study of angular distributions
(as in Ref.~\cite{Bruser:2015yka} for the case of polarised
$\tau$-lepton decays), the charges of the external states need not be
specified. The following partial widths should be divided by the total
muon decay width, \emph{i.e.}
$\Gamma_\mu\simeq\left(G_F^2m_\mu^5\right)/\left(192\pi^3\right)$, in
order to obtain the corresponding BRs.

\vspace{1.2cm}
\begin{boldmath}
\subsection{$\mu\to e \gamma$}
\end{boldmath}
\noindent
The {simplest and most investigated} LFV muonic process is $\mu\to e\gamma$. On the
one hand, the serious experimental bounds~\cite{TheMEG:2016wtm} on
this kinematically simple transition clearly indicate that there is an
indisputable conservation mechanism at work. On the other hand, any
observation of a non-zero $\mu\to e\gamma$ in current or future
experiments would indicate the existence of BSM physics. The Lagrangian in Eq.~\ref{Leff} results in a branching ratio
\begin{align}
\label{muegBR}
{\Gamma}\left( {\mu  \to e\gamma } \right)  =  
\frac{e^2 m_\mu ^5}{{4\pi \Lambda^4 }}
   \left( {{{\left| {C^{D}_L} \right|}^2} 
+ {{\left| {C^{D}_R} \right|}^2}} \right)\,,
\end{align}
from which it is clear that, {with the Wilson coefficients defined at the muon energy scale}, the associated
BR is related only to the dipole operators $C^{D}_{L/R}$.  According
to the RGEs presented in Eq.~\ref{rgedipole}, these operators will
receive contributions from scalar ($C^S_{ll}$ with $l=e,\mu$) and
tensor ($C^T_{\tau\tau}$ and $C^T_{qq}$ with $q=u,c,d,s,b$) operators,
with non-vanishing coefficients at higher scales.

\vspace{1.2cm}
\begin{boldmath}
\subsection{$\mu\to eee$}
\end{boldmath}
\noindent
The second representative channel for muonic LFV decays is $\mu\to
eee$. Prospects for future experimental developments in this rare muon
process are very promising: the current experimental
limit~\cite{Bellgardt:1987du} is expected to be improved considerably
by the Mu3e experiment. Again, any signal of such a rare decay would be
a clear signal for BSM physics.

The partial width reads
\begin{align}
\label{mu3eBR}
&\Gamma\left(\mu  \to 3e\right)=\nonumber \\
&=\frac{\alpha^2 m_\mu^5 }{12 \Lambda^4 \pi }\left(\left|C^{D}_{L}\right|^2+\left|C^{D}_{R}\right|^2\right) \left(8 \log\left[\frac{m_\mu}{m_e}\right]-11\right)\nonumber\\
&+\frac{m_\mu^5}{3 \Lambda^4 (16\pi)^3} \left(\left|C_{ee}^{S\;LL}\right|^2+\left|C^{S\;RR}_{ee}\right|^2+8 \left(2 \left|C_{ee}^{V\;LL}\right|^2+\left|C_{ee}^{V\;LR}\right|^2+\left|C_{ee}^{V\;RL}\right|^2+2 \left|C_{ee}^{V\;RR}\right|^2\right)\right)
\nonumber \\
&-\frac{\alpha m_\mu^5 }{3 \Lambda^4 (4\pi)^2}(\Re[C^{D}_{L} \left(C_{ee}^{V\;RL}+2C_{ee}^{V\;RR}\right)^*]+\Re[C^{D}_{R} \left(2C_{ee}^{V\;LL}+C_{ee}^{V\;LR}\right)^*]),
\end{align}
where a more complicated {interplay} between operators occurs. The next section provides an
explicit example of a correlation between the coefficients in
Eqs.~\ref{muegBR}~and~\ref{mu3eBR} with respect to the two
experimental bounds on LFV transitions.

\vspace{1.2cm}
\begin{boldmath}                                            %
\section{Limits on Wilson coefficients and correlations}
\label{sec:4}
\end{boldmath}                                              %
\noindent
In this section, the present experimental limits together with
anticipated updates are applied to the {observables} of
Eqs.~\ref{muegBR}~and~\ref{mu3eBR} defined at a UV-completion energy
scale. 

Closer examination of Eqs.~\ref{muegBR}~and~\ref{mu3eBR} together with
the RGE equations in the appendix reveals that only two classes of
operators -- the dipole ($O^D$) and the scalar ($O^{S}_{ee}$) -- are
manifestly correlated at the one-loop level in two self-consistent
systems (separate by chirality) of ordinary differential equations
(ODE). In principle, more complicated relations occur if non-zero
tensorial quark or $\tau$-lepton operators are considered. In
addition, at the two-loop level, even the vectorial operators mix with
the dipole. However, a complete quantitative treatment of all possible
correlations is beyond the scope of this analysis.

For illustrative purposes, in the following discussion, we consider a
scenario where an underlying UV-complete theory produces non-vanishing
SMEFT coefficients.  We assume that matching this SMEFT to the
low-energy Lagrangian of Eq.~\ref{Leff}, only two categories of
non-vanishing coefficients are produced, namely $C^D$ and
$C^{S}_{ee}$.

According to the RGE described by
Eqs.~\ref{rgedipole}~and~\ref{rgescalar}, if {the RGE effects are neglected for} the EM coupling and
fermion masses\footnote{If the running
  of the electromagnetic (EM) coupling and the fermion masses is taken
  into account, then the evolution of the couplings is more involved,
  but at the same time the qualitative conclusion of this note will
  remain unchanged.}, then the running of these two operators can be
described by a relatively simple system of two ODE. The solutions are
\begin{align}\label{eq:diff1}
C^D_{L/R}\left(\mu\right)&\simeq
\left(\frac{\mu}{m_Z}\right)^{4 \widetilde{\alpha}}C^D_{L/R}\left(m_Z\right)
-\frac{ m_e }{16 \alpha \pi m_\mu }\left(\frac{\mu}{m_Z}\right)^{3 \widetilde{\alpha}}\left(\frac{m_Z^{\widetilde{\alpha}}-\mu^{\widetilde{\alpha}}}{m_Z^{\widetilde{\alpha}}}\right)
C^{S\; LL/RR}_{ee}\left(m_Z\right),\\\label{eq:diff2}
C^{S\; LL/RR}_{ee}\left(\mu\right)&\simeq\left(\frac{\mu}{m_Z}\right)^{3 \widetilde{\alpha}}C^{S\; LL/RR}_{ee}\left(m_Z\right),
\end{align}
where $\mu$ is the phenomenological energy scale at which the
coefficients should be evaluated, and $\widetilde{\alpha}=\alpha/\pi$
is the normalised EM coupling. 

By combining these results with the BRs of Section~\ref{sec:3} and
applying the experimental limits, at the muon mass scale $\mu=m_\mu$,
we obtain the constraints on the coefficients $C^D(M_Z)$ and
$C^{S}_{ee}(M_Z)$ shown in Figure~1 (right-chirality ones give the
same result).  Note that  the evolution of the EM coupling and fermion
masses is taken into account in these numerical results.
\begin{figure}[!th]
\begin{minipage}{\linewidth}
\centerline{\includegraphics[width=0.89\linewidth]{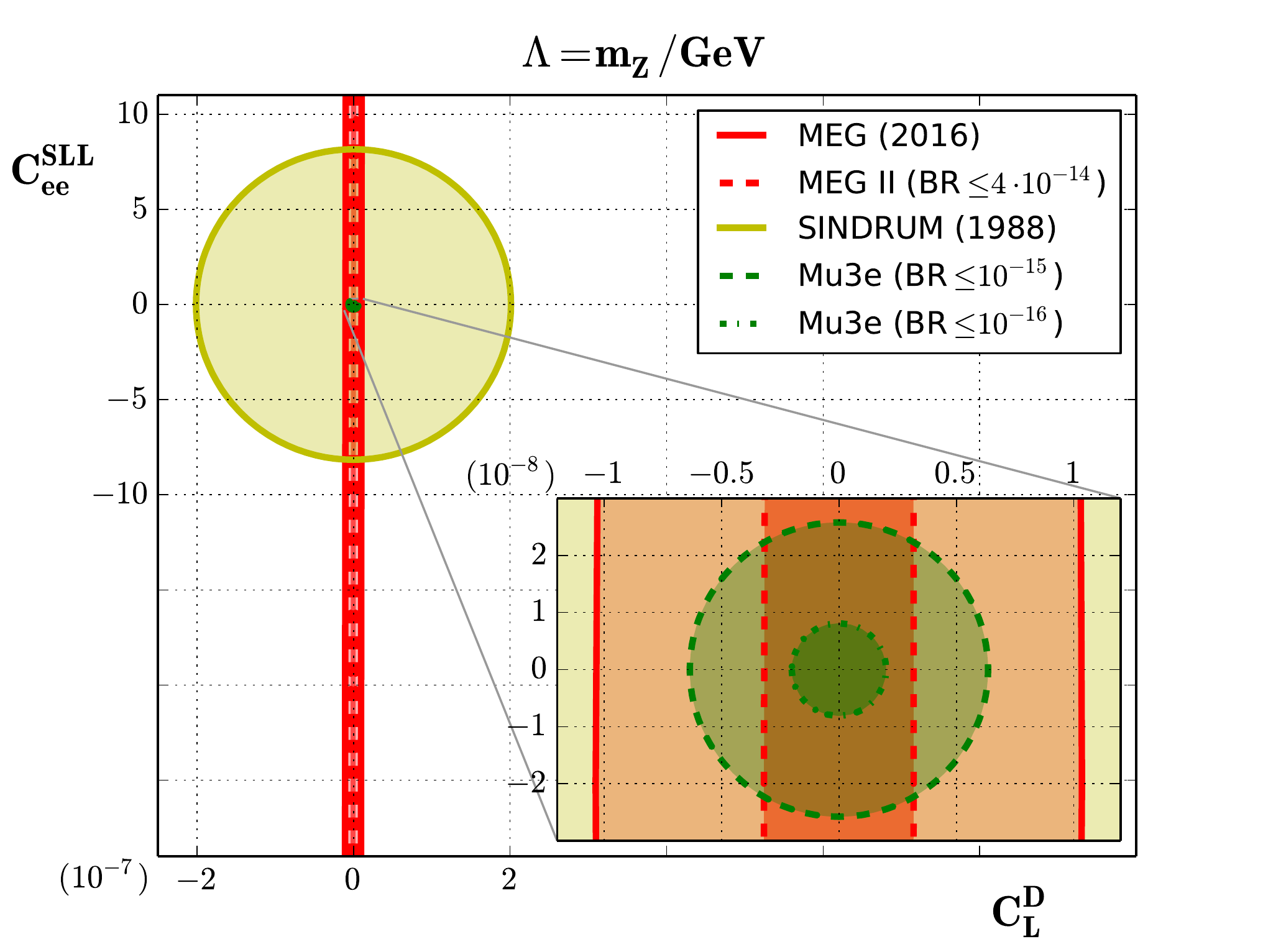}}
\end{minipage}
\begin{minipage}{\linewidth}
\centerline{\includegraphics[width=0.89\linewidth]{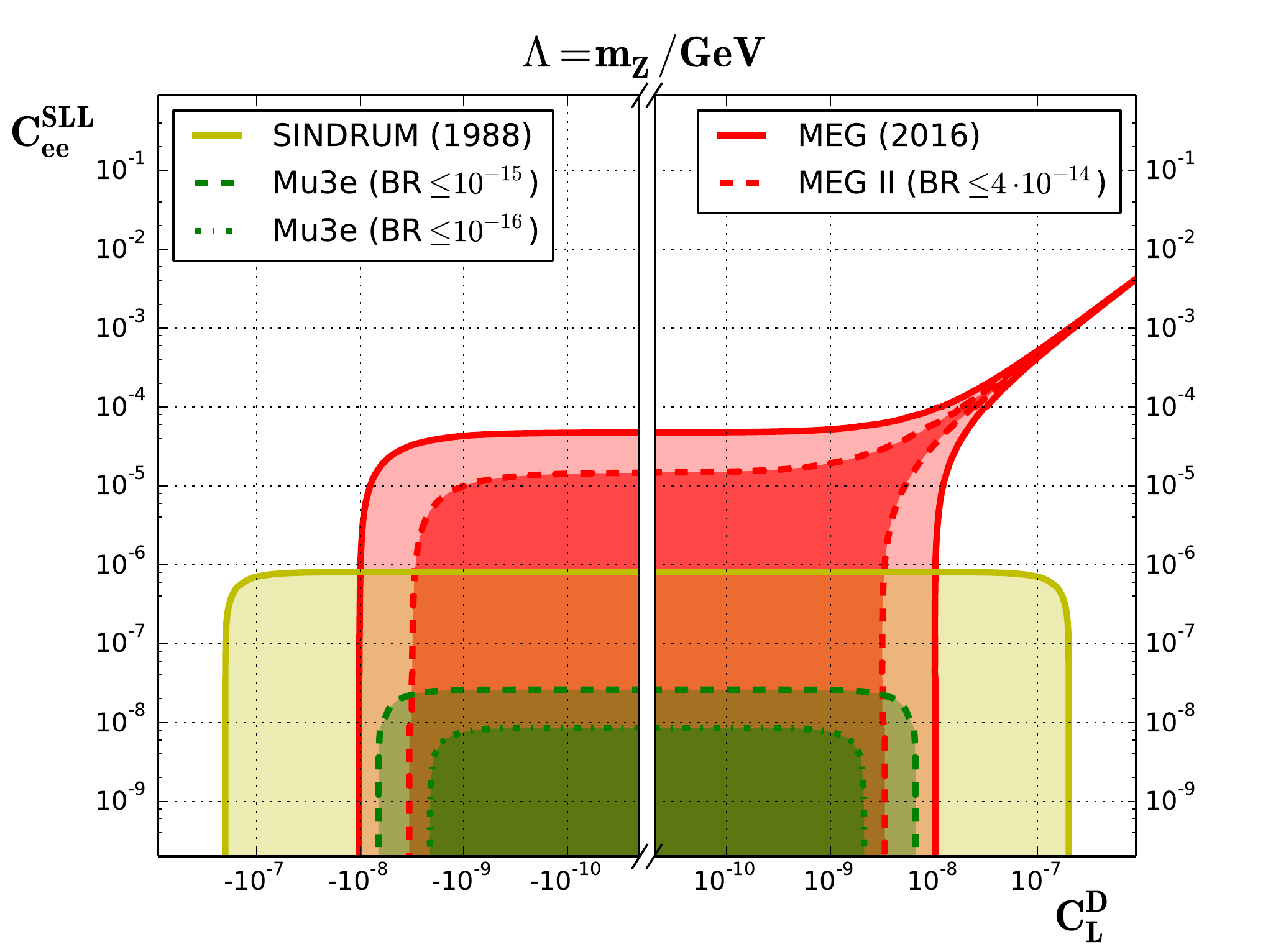}}
\end{minipage}
\caption{Allowed parameter space for the two coefficients $C^{D}_{L}$
  and $C^{S\; LL}_{ee}$ (non-vanishing at the $\Lambda=m_Z$ energy
  scale) by the $\mu\to e\gamma$ (red regions) and $\mu \to 3e$
  experiments (yellow/green regions). Present (solid lines) and
  anticipated bounds (dashed/dashed-dotted lines) are plotted on a
  linear (upper frame) and pseudo-logarithmic scale (lower frame). In
  evaluating of the RGE, the running of the gauge coupling and fermion
  masses are included.} 
\end{figure}


First, it must be appreciated that the limits originating from the
non-observation of LFV muon decays in different experiments are
manifestly complementary. In particular, for $\mu\to e\gamma$ there is
a region of the parameter space in which an explicit cancellation
occurs between the contributions of the two operators. This effect is
due to the relative sign in the evolution equation, which implies that
$C^{D}_{L/R}(m_\mu)$ is small if
\begin{align}
C^D_{L/R}\left(m_Z\right)\simeq
\frac{ m_e }{16 \alpha \pi m_\mu }\left(\frac{m_Z^{\widetilde{\alpha}}
-m_\mu^{\widetilde{\alpha}}}{m_\mu^{\widetilde{\alpha}}}\right)
C^{S\; LL/RR}_{ee}\left(m_Z\right).
\end{align}
Thus for MEG there is a blind direction in parameter space. In
contrast, the $\mu\to 3e$ decay mode is not subject to any
cancellation among effective couplings, meaning that only the
future Mu3e experiment will be able to explore this corner of the
parameter space, as the SINDRUM experiment did in the past.

A second important aspect is that the last stage of the Mu3e
experiment will cover a wider region of the parameter space than the
MEG II experiment (in the absence of other correlations between
operators), producing better limits for both the dipole and
four-fermion effective couplings. 

A much more involved scenario might arise if other operators are taken
into account. For example, if $C^{T}_{bb}$ is generated at the EWSB
energy scale, the evolution of the dipole operator changes
dramatically. However, salient aspects of the complementarity of the
two experimental searches will remain qualitatively unaltered.

\vspace{1.2cm}
\begin{boldmath}                                            %
\section{Conclusion}
\label{sec:5}
\end{boldmath}                                              %
\noindent
In this note, LFV muon decays have been analysed within the framework
of an effective field theory with higher-dimensional operators at low
energy scales.

The processes $\mu\to e \gamma$ and $\mu\to 3e$ have been
investigated below the EWSB energy scale, down to the natural energy
regime at which such processes occur, \emph{i.e.} the muon mass
scale. The complete class of contributing dimension-5 and dimension-6
operators allowed by QED and QCD have been systematically studied at
the tree level, and one-loop contributions to the RGE have been taken
into account.

The current experimental limits from the MEG and SINDRUM experiments
have been used to extract bounds on some of the Wilson coefficients of
the effective theory and, ultimately, on the Wilson coefficients at
any energy level below the EWSB scale.

This note has also presented an explicit example of a correlation
between dipole and four-fermion scalar effective couplings, under the
assumption that they are the only two non-vanishing couplings
generated at the EWSB energy scale by an underlying BSM theory,
illustrating the complementarity of the searches planned for the
MEG~II and Mu3e experiments. In particular, it has been shown that the
$\mu\to 3e$ channel allows for exploration of a region of the
parameter space which $\mu\to e\gamma$ experiments are unable to
investigate. Furthermore, in the absence of any other correlation it
was shown that the last experimental phase of Mu3e will provide the
best bound on the parameter space for both considered
operators. However, this assertion might be invalid in the presence of
other operators that mix in some way with the tree-level Wilson
coefficients.
%

\vspace{1.2cm}
\begin{boldmath}                                            %
\section*{Acknowledgements}
\end{boldmath}                                              %
\noindent
AC's work is supported by an Ambizione grant from the Swiss National
Science Foundation (SNF).  GMP's work is supported by SNF under
contract 200021\_160156.  GMP thanks Angela Papa and Ann-Kathrin
Perrevoort for their insight and expertise, which greatly improved the
manuscript.

\vspace{1.2cm}
\begin{boldmath}                                            %
\section*{Appendix - Anomalous dimensions}
\label{app:A}
\end{boldmath}                                              %
\noindent
This appendix presents the anomalous dimensions of the operators
exploited in the phenomenological analysis of Section~\ref{sec:4}. The
corresponding equations for the chirality-flipped operators are
obtained by the label interchange $R\longleftrightarrow L$.

The dipole operator runs according to
\begin{align}\label{rgedipole}
16\pi^2\frac{\partial C^{D}_{L}}{\partial \left(\log{\mu}\right)}&=
16 e^2 Q_l^2 C^{D}_{L}\nonumber\\
&-Q_l\frac{m_{e}}{m_{\mu}}C^{S\;LL}_{ee}
-Q_l C^{S\;LL}_{\mu\mu}
+8Q_l\frac{ m_{\tau} }{m_{\mu}}C^{T\;LL}_{\tau\tau}\Theta(\mu-m_\tau)
\nonumber\\
&+\frac{8N_c}{m_{\mu}}\sum_{q}m_{q}Q_qC^{T\;LL}_{qq}\Theta(\mu-m_q).
\end{align}
where $N_c$ is the number of colours, and $Q_l$, $Q_u$ and $Q_d$ are
the charges associated with leptons, $u$-type and $d$-type quarks,
respectively. 

The running of the leptonic scalar and tensorial operators is
summarised by the following equations: 
\begin{align}\label{rgescalar}
16\pi^2\frac{\partial C^{S\;RR}_{ee/\mu\mu}}{\partial \left(\log{\mu}\right)}=12e^2Q_l^2 C^{S\;RR}_{ee/\mu\mu},
\end{align}
\begin{align}
16\pi^2\frac{\partial C^{S\;RR}_{\tau\tau}}{\partial \left(\log{\mu}\right)}=
-12e^2Q_l^2 
\left(
C^{S\;RR}_{\tau\tau}+8C^{T\;RR}_{\tau\tau}
\right),
\end{align}
\begin{align}
16\pi^2\frac{\partial C^{S\;RL}_{\tau\tau}}{\partial \left(\log{\mu}\right)}=
-12e^2Q_l^2 C^{S\;RL}_{\tau\tau},
\end{align}
\begin{align}
16\pi^2\frac{\partial C^{T\;RR}_{\tau\tau}}{\partial \left(\log{\mu}\right)}=
-2e^2Q_l^2 
\left(
C^{S\;RR}_{\tau\tau}-2C^{T\;RR}_{\tau\tau}
\right).
\end{align}

The running of the scalar and tensorial quark operators is given by
\begin{align}
16\pi^2\frac{\partial C^{S\;RR}_{qq}}{\partial \left(\log{\mu}\right)}=
\left(-6\left(Q_l^2+Q_q^2\right)e^2
+\left(1-N_c^2\right)g_S^2\right)C^{S\;RR}_{qq}
-96e^2Q_l Q_q C^{T\;RR}_{qq},
\end{align}
and
\begin{align}
16\pi^2\frac{\partial C^{T\;RR}_{qq}}{\partial \left(\log{\mu}\right)}=
-2e^2Q_l Q_qC^{S\;RR}_{qq}
+\left(2\left(Q_l^2+Q_q^2\right)e^2+\left(\frac{N_c^2-1}{N_c}\right)g_S^2\right)C^{T\;RR}_{qq}.
\end{align}

The running of vector operators is decoupled from the dipole operator
$C^D$ at the one-loop level. Nevertheless, it is well known that a
non-vanishing mixing occurs at the two-loop
level~\cite{Ciuchini:1993ks,Ciuchini:1993fk}. However,
inclusion of these effects is beyond the scope of the present analysis
and will be provided in a future publication~\cite{future}.

\vspace{1.2cm}
\section*{References}

\bibliography{clfv}






\end{document}